\documentclass[twoside,a4paper,11pt]{sea10}
\usepackage{graphicx}
\begin{document}
\pagenumbering{arabic}
\pagestyle{myheadings}
\thispagestyle{empty}
{\flushleft\includegraphics[width=\textwidth,bb=58 650 590 680]{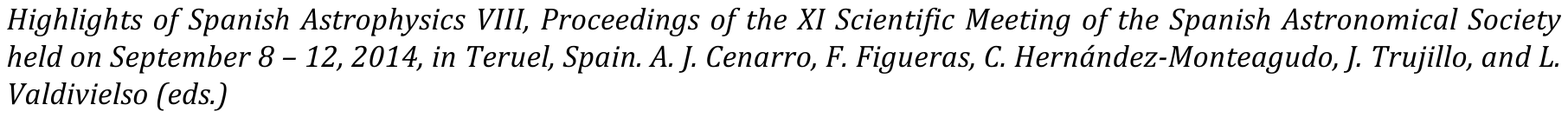}}
\vspace*{0.2cm}
\begin{flushleft}
{\bf {\LARGE
%
Origin of Galactic Type-Ia supernovae: SN 1572 and SN 1006
%
}\\
\vspace*{1cm}
%
Jonay I. Gonz\'alez Hern\'andez$^{1,2}$,
Pilar Ruiz-Lapuente$^{3,4}$, 
Hugo M. Tabernero$^{5}$, 
David Montes$^{5}$, 
Ram\'on Canal$^{3}$, 
Javier M\'endez$^{6}$, 
and 
Luigi R. Bedin$^{7}$
%
}\\
\vspace*{0.5cm}
%
$^{1}$
Instituto de Astrof{\'\i}sica de Canarias, E-38205 La Laguna, Tenerife, Spain\\
$^{2}$
Universidad de La Laguna, Dpto. de Astrof{\'\i}sica, E-38206 La Laguna, Tenerife, Spain\\
$^{3}$
Department of Astronomy, Institut de Ci\`encies del Cosmos, Universitat de Barcelona
(UB-IEEC), Mart{\'\i} i Franqu\'es 1, E-08028 Barcelona, Spain\\
$^{4}$
Instituto de F{\'\i}sica Fundamental, Consejo Superior de Investigaciones Cient{\'\i}ficas, 
Serrano 121, E-28006, Madrid, Spain\\
$^{5}$
Departamento  de Astrof{\'\i}sica y Ciencias de la Atm\' osfera, Facultad de Ciencias Físicas, 
Universidad Complutense de Madrid, E-28040 Madrid, Spain\\
$^{6}$
Isaac Newton Group of Telescopes, P.O. Box 321; E-38700 Santa 
Cruz de La Palma, Spain\\
$^{7}$
INAF-Osservatorio Astronomico di Padova, Vicolo dell'Osservatorio 5,
I-35122 Padova, Italy\\
%
\end{flushleft}
%
\markboth{
Origin of SN 1572 and SN 1006
}{ 
%
Gonz\'alez Hern\'andez et al. 
%
}
\thispagestyle{empty}
\vspace*{0.4cm}
\begin{minipage}[l]{0.09\textwidth}
\ 
\end{minipage}
\begin{minipage}[r]{0.9\textwidth}
\vspace{1cm}
\section*{Abstract}{\small
%
We have been searching for surviving companions of progenitors of Galactic Type-Ia supernovae, in particular SN 1572 and SN 1006. These companion stars are expected to show peculiarities: (i) to be probably more luminous than the Sun, (ii) to have high radial velocity and proper motion, (iii) to be possibly enriched in metals from the SNIa ejecta, and (iv) to be located at the distance of the SNIa remnant. We have been characterizing possible candidate stars using high-resolution spectroscopic data taken at 10m-Keck and 8.2m-VLT facilities. We have identified a very promising candidate companion (Tycho G) for SN 1572 (see \cite{rui04,gon09a,bed14}, however for a different view see \cite{ker12}) but we have not found any candidate companion for SN 1006, suggesting that SN event occurred in 1006 could have been the result of the merging of two white dwarfs (see \cite{gon12}). 
Adding these results to the evidence from the other direct searches, the clear minority of cases (20\% or less) seem to disfavour the single-degenerate channel or that preferentially the single-degenerate escenario would involve
 main-sequence companions less massive than the Sun. Therefore, it appears to be very important to continue investigating these and other Galactic Type-Ia SNe such as the Johannes Kepler SN 1604.
%
\normalsize}
\end{minipage}
%
%
%
\section{Introduction \label{intro}}
Type Ia supernovae (SNe Ia) are the best known cosmological distance indicators at high redshifts. Their use led to the discovery of the currently accelerating expansion of the universe (see \cite{riess98}, \cite{per99}). These SNe are thought to occur when a white dwarf made of carbon and oxygen accretes sufficient mass to trigger a thermonuclear explosion. The explosion could happen via accretion from a companion star (single degenerate (SD) channel) or via merging of two white dwarfs (double-degenerate (DD) channel). Therefore, a companion star will survive the explosion only in the SD channel (see these reviews \cite{how11,wan12,mao14,rui14}, for further information). Both channels might contribute to the production of Type-Ia supernovae but the relative proportions of their contributions remain unclear.

One way to investigate Type-Ia SNe is by performing direct survey of the field of historical events~\cite{rui97}. Depending on the date of the SN explosion and the distance to the SN remnant, we are able to define the region where to find the possible companion star, close to the geometrical center of the SN remnant~\cite{rui14}.

There are a few known historical Galactic Type-Ia supernova events: SN 1572 (Tycho Brahe supernova), SN 1006, SN 1604 (Kepler supernova), and three new recently identified Type-Ia supernova remnants: SN 185 (RCW86), G299.2-2.9 and G272.2-3.2. Our group have been investigating these Galactic historical SNIa, trying to search for companion stars of progenitors of historical Galactic Type-Ia SNe with the aim of clarifying the origin of these cosmological candles. We present here the results we have got in two of them, SN 1572 and SN 1006. 

\section{SN 1572: Tycho Brahe supernova}

Ruiz-Lapuente et al. \cite{rui04} investigated the Type-Ia SN 1572 and found one promising candidate (called Tycho G) to be the surviving companion of the progenitor of SN 1572 remnant, thus the origin of Tycho's supernova remnant (SNR)
might be attributed to the SD channel. On the other hand, for instance, the absence of any ex-companion in the supernova remnant SNR 0509$-$67.5, in the Large Magellanic Cloud (LMC), down to very faint magnitudes, strongly suggests that the SN explosion there was produced by a DD system.
 
\begin{figure}
\center
\includegraphics[width=7.1cm,angle=0,clip=true]{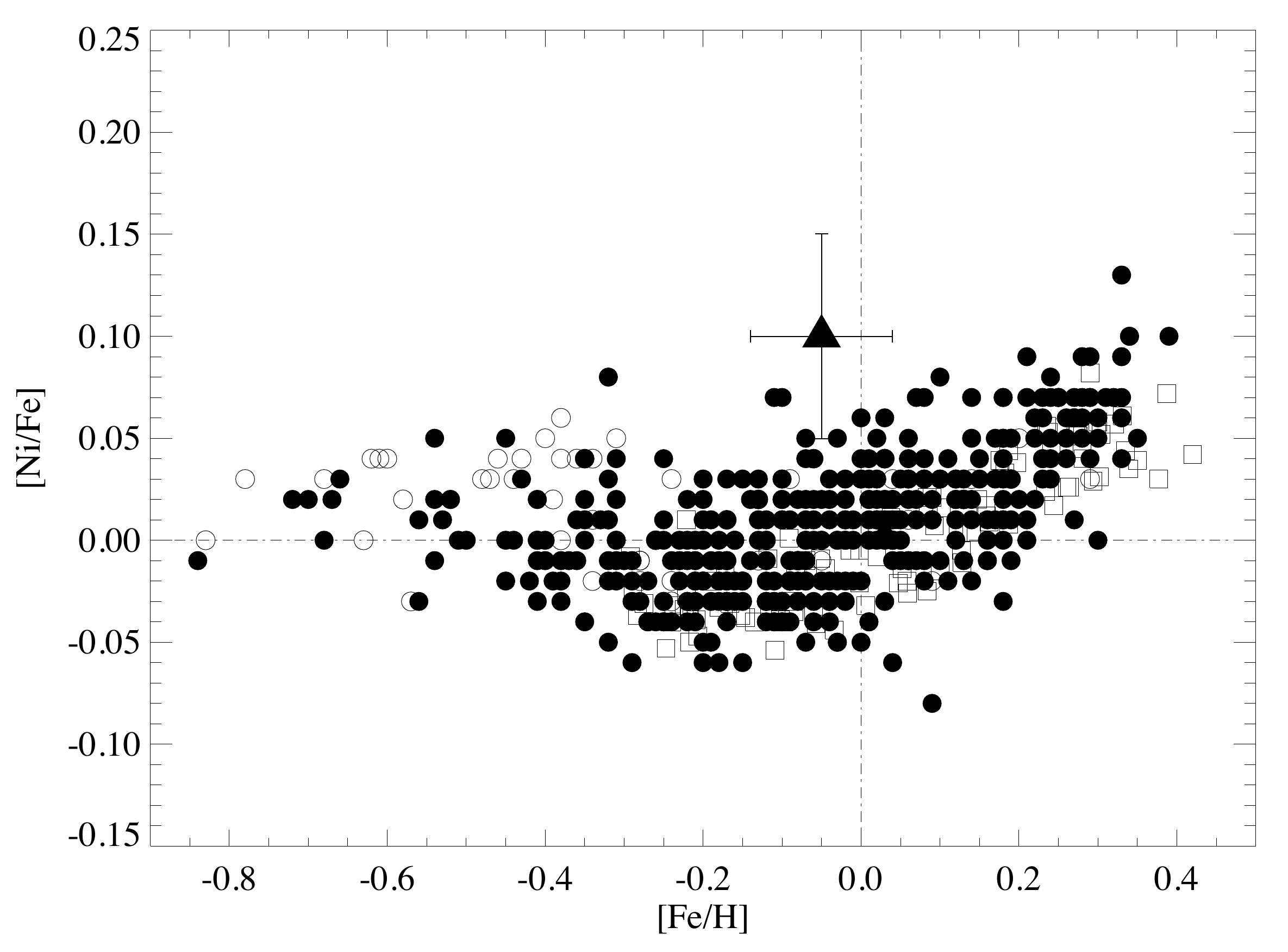} 
\includegraphics[width=7.1cm,angle=0,clip=true]{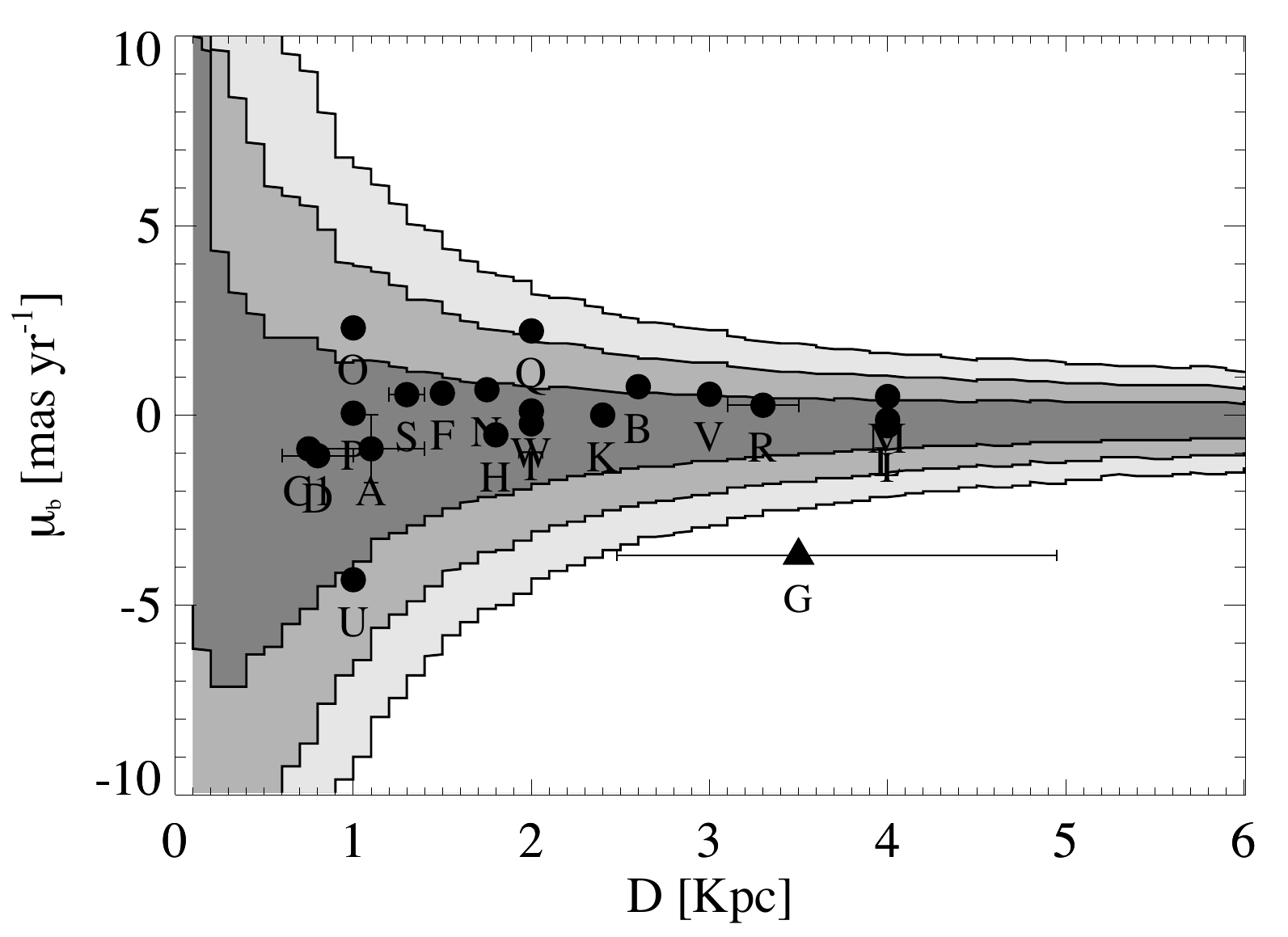} 
\caption{\label{fig1} 
\emph{Left panel:} [Ni/Fe] abundance ratio of Tycho-G (filled triangle) in comparison with the abundances of F-, G- and K-type metal-rich dwarf stars from \cite{nev09}. Thin-disc stars are depicted as filled circles, whereas transition and thick-disc stars are the empty circles. Solar analogues are shown as empty squares from \cite{gon10, gon13}. The size of the error bars indicates the 1$\sigma$ uncertainty. The dashed-dotted lines indicate solar abundance values. 
\emph{Right panel:} 
The distributionof proper motions, $\mu_b$, perpendicular to the
Galactic plane, as a function of distance, in the direction of the
X-ray centroid of Tycho's SNR, for thin-disc and thick-disc stars
together (and [Fe/H]~$ > -0.14$), according to the Besan\c con model
of the Galaxy (see~\cite{rob03}). 1-$\sigma$, 2-$\sigma$ and 3-$\sigma$ 
regions are indicated. The position of Tycho-G is depicted as a 
triangle, $D$[kpc]~$= 3.50\pm1.45$ and $\mu_b$[mas yr$^{-1}$]~$ 
= -3.69 \pm 0.10(0.04)$.
}
\end{figure}

Gonz\'alez Hern\'andez et al. \cite{gon09a} studied the star Tycho G and derived its stellar parameters and chemical abundances, using HIRES@10m-KeckI high-resolution spectra. In Fig.~\ref{fig1}, we show a re-determination of the Ni abundance in Tycho G compared with those of similar stars of the Galactic thin and thick disk. A slightly Ni enhancement is seen in Tycho G (see \cite{gon09a,bed14}), which may suggest  Ni pollution from Tycho SN 1572 ejecta. We have been also improved our previous proper motion determination of Tycho G together with many other stars in the Tycho SN field using images from HST programmes GO-9729, GO-10098 and GO-12469 (PI: Ruiz-Lapuente), which span about 8 yr. In Fig.~\ref{fig1} we compare the proper motion (PM) of Tycho G, perpendicular to the Galactic plane, together with those of other stars of the SN 1572 field, in comparison with the distribution of PMs predicted by the Besan\c con model of the Galaxy \cite{rob03}. Tycho G is the only peculiar star in Fig.~\ref{fig1}, and thus possibly the only viable candidate to be the surviving companion of SN 1572~\cite{bed14}.
Several models have been proposed to evaluate the luminosity, peculiar radial velocity and proper motion, and rotational velocity of Tycho G and support this star as the possible surviving companion~\cite{pan14}.

\section{SN 1006: The Galactic brightest apparent stellar event}

\begin{figure}
\center
\includegraphics[width=10.1cm,angle=0,clip=true]{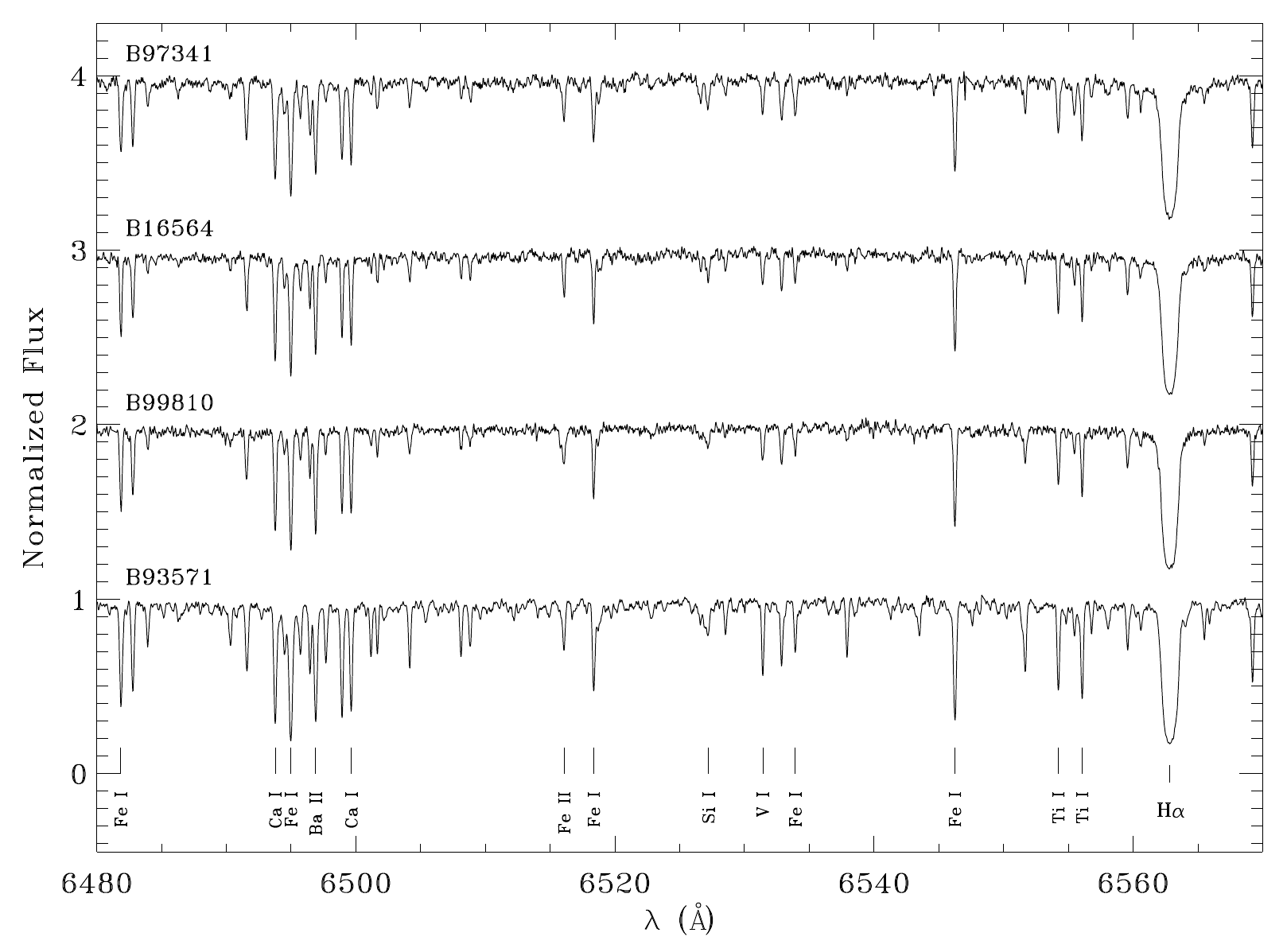} 
\caption{\label{fig2} Small portion of the high-resolution UVES spectra of giant stars at the distance of the SN 1006 remnant sorted in decreasing effective temperature order from top to bottom. 
}
\end{figure}

\begin{figure}
\center
\includegraphics[width=10.1cm,angle=0,clip=true]{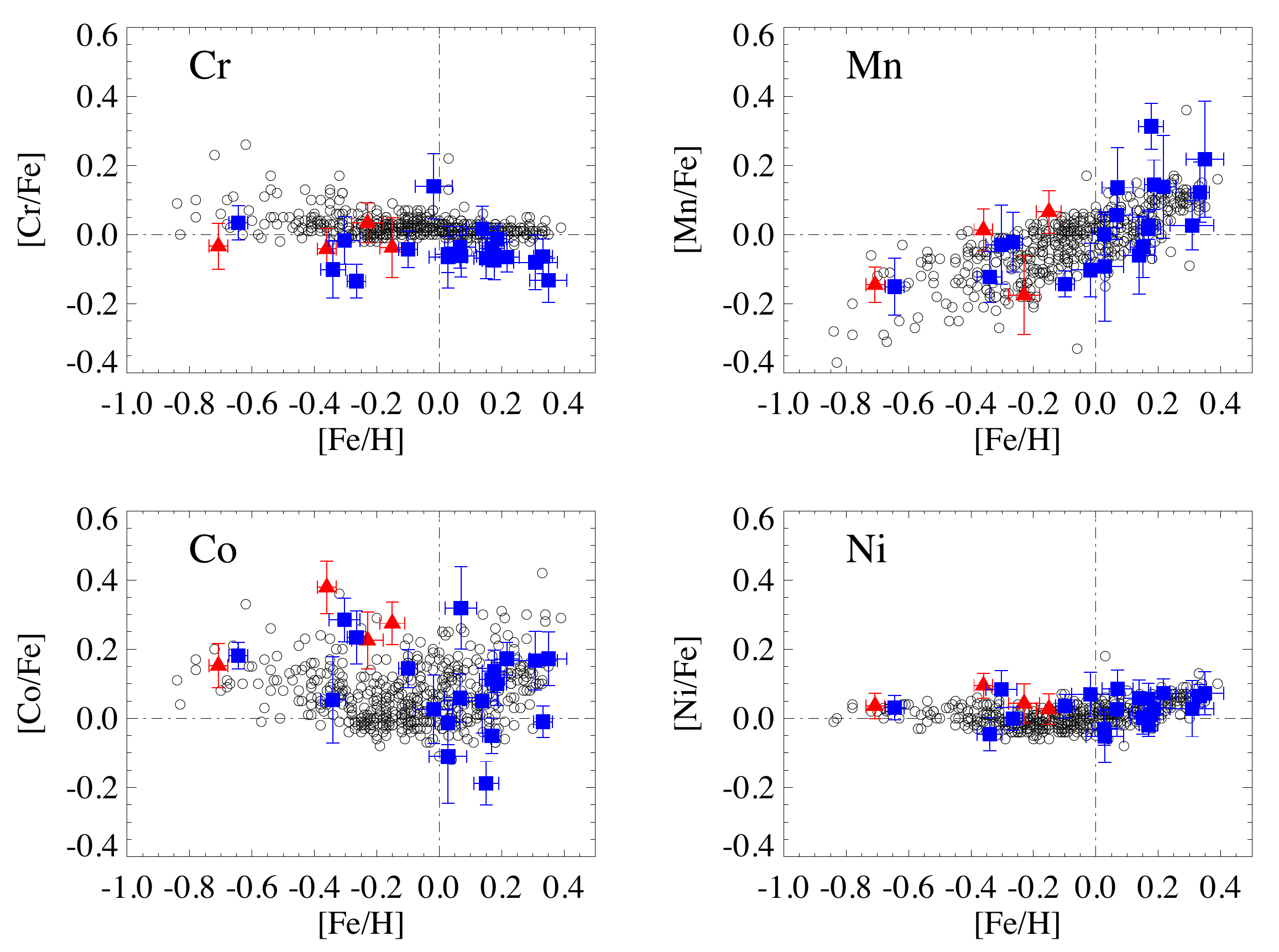} 
\caption{\label{fig3} Chemical abundances of Fe-peak elements derived using the EW technique. We performed a differential analysis on a line-by-line basis, using the solar UVES spectrum of the Moon as reference. Red triangles correspond to the four giant stars whose distances are marginally compatible with that of the remnant of SN 1006. Blue squares, to the rest of the stars in the sample. 
}
\end{figure}

The SN 1006 has been also investigated with the aim of finding the surviving companion of the SN progenitor (see \cite{gon12, ker12}). We selected all targets up to a limiting magnitude of $m_R=15$, and we acquired high-resolution spectroscopy using UVES@8.2m-VLT within a circle of radius 4’ around the geometrical centre of the SN 1006 remnant. In Fig.~\ref{fig2} we show UVES@VLT observed spectra of several giant stars located at distances consistent with the SN 1006 remnant. Model simulations of the impact of SNIa ejecta indicate that red giant should have been stripped off most of its hydrogen envelope and therefore, these stars, which are normal giant stars, are not expected to be the surviving companion. The stellar parameters were derived from the measurements of equivalent widths of FeI-II lines using the code StePar~\cite{tab12}.
In Fig.~\ref{fig3} we depict the abundance ratios [X/Fe] of several Fe-peak elements of the stars in the SN 1006 field compared to those in stars belonging to the Galactic disk. None of the stars reveals any remarkable abundance peculiarity. 

\begin{figure}
\center
\includegraphics[width=10.1cm,angle=0,clip=true]{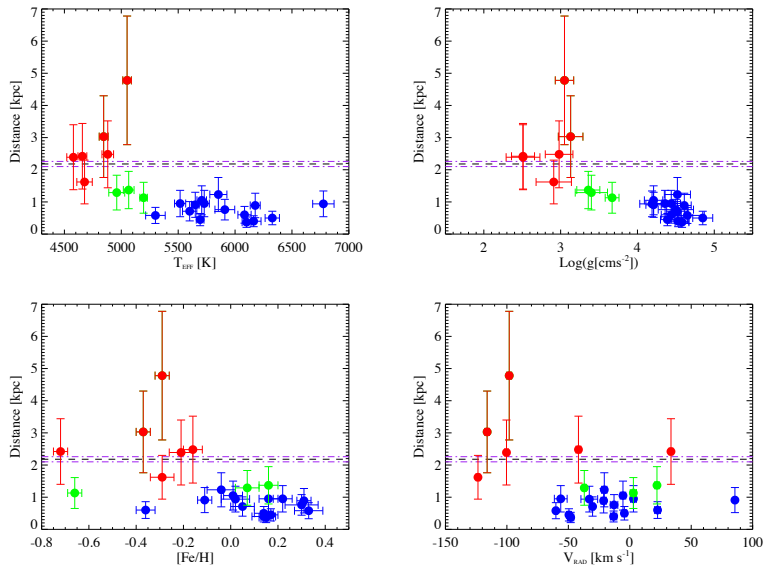} 
\caption{\label{fig4} Distances of the target stars within the SN 1006 field as a function of their stellar parameters, metallicity and radial velocity. Giant, subgiant and dwarf stars are shown as red, green and blue filled circles, respectively. The black dashed line gives the distance of the SN 1006 remnant. The violet dashed-dotted line show the well-defined distance error range of the SN 1006 remnant. 
}
\end{figure}

Using optical and near-infrared photometry together with the stellar parameters we derive the distances to the stars of the SN 1006 field (see Fig.~\ref{fig4}). We clearly see that only giant stars are compatible within the error bars with the distance to the SN remnant. We inspect the 2MASS photometry catalogue and use the IRFM method (see \cite{gon09b}) to derive effective temperatures of the additional candidates, with the aim of identifying those at the distance of the SN 1006 remnant. We found no main-sequence brighter than $m_R \sim 16.4$, which brings the limit down to $M_R \sim +4.5$, corresponding to $M_V \sim +4.9$ (approximately equal to, or slightly less than, solar luminosity). Hydrodynamical simulations suggest that even if the pre-SN companion is main-sequence solar-mass star, about 1000 yr after the strong impact of the SNIa ejecta, the star does not have enough time to become dimmer than the Sun. However, Di Stefano, Voss and Claeys (\cite{diste}) have suggested that, by the time that the exploding white dwarf has reached the critical mass for explosion (within the single degenerate scneario), the companion (donor) might have become a WD. This finding deserves further investigation for its implications in the surviving companion of SNe Ia.

Several stars analyzed in \cite{gon12} are compared to model predictions, but those models gives hotter candidates than those observed (see \cite{pan14} for further details). Thus, SN 1006, the brightest event ever observed in our Galaxy, should have been possibly produced by merging of two white dwarfs. Adding this result to the evidence from the other direct searches, the single-degenerate channel appears either to happen in the minority of the cases (20\% or less), or preferentially it involves main-sequence companions with masses more probably below that of the Sun.
%
%
\small  
%
\section*{Acknowledgments}   
%
J.I.G.H. acknowledges financial support from the Spanish Ministry projects MINECO AYA2011-29060 and MINECO AYA2011-26244, and also from the Spanish Ministry of Economy and Competitiveness (MINECO) under the 2011 Severo Ochoa Program MINECO SEV-2011-0187. P.R.L. is grateful for the support of the Spanish Ministry project MINECO AYA2012–36353. H.M.T and D.M acknowledge financial support from the Universidad Complutense de Madrid (UCM), the Spanish Ministry of Econ- omy and Competitiveness (MINECO) from pojects AYA2011-30147-C03-02, and The Comunidad de Madrid under PRICIT project S2009/ESP-1496 (Astro- Madrid). H.M.T also acknowledges the financial support of the Spanish Ministry of Economy and Competitiveness (MINECO) under grants BES-2009-012182 and EEBB-I-12-04038.
%

%
\end{document}